\documentclass[fleqn,usenatbib]{mnras}

\usepackage{newtxtext,newtxmath}

\usepackage[T1]{fontenc}

\DeclareRobustCommand{\VAN}[3]{#2}
\let\VANthebibliography\thebibliography
\def\thebibliography{\DeclareRobustCommand{\VAN}[3]{##3}\VANthebibliography}

\usepackage{graphicx}	

\def\dg{\hbox{$^\circ$}}

\title[Chaos in \textit{Fermi}-LAT blazars]{Searching for signatures of chaos in $\gamma$-ray light curves of selected \textit{Fermi}-LAT blazars}

\author[O. Ostapenko et al.]{ 
O. Ostapenko,$^{1}$\thanks{E-mail: sash.stesnyash@knu.ua}
M. Tarnopolski,$^{2}$\thanks{E-mail: mariusz.tarnopolski@uj.edu.pl}
N. \.{Z}ywucka$^{3}$\thanks{E-mail:n.zywucka@oa.uj.edu.pl}
and J. Pascual-Granado$^{4}$\thanks{E-mail: javier@iaa.es}
\\
$^{1}$Department of Astronomy and Space Physics, Taras Shevchenko National University of Kyiv, Akademika Hlushkova Ave 4, Kyiv, 03680, Ukraine\\
$^{2}$Astronomical Observatory, Jagiellonian University, Orla 171, 30-244, Krak\'ow, Poland\\
$^{3}$Centre of Space Research, North-West University, Potchefstroom, South Africa\\
$^{4}$Instituto de Astrof\'isica de Andaluc\'ia (IAA-CSIC), Glorieta de la Astronomía s/n, 18008, Granada, Spain\\
}

\date{Accepted XXX. Received YYY; in original form ZZZ}

\pubyear{2020}

\begin{document}
\label{firstpage}
\pagerange{\pageref{firstpage}--\pageref{lastpage}}
\maketitle

\begin{abstract}
Blazar variability appears to be stochastic in nature. However, a possibility of low-dimensional chaos was considered in the past, but with no unambiguous detection so far. If present, it would constrain the emission mechanism by suggesting an underlying dynamical system. We rigorously searched for signatures of chaos in \textit{Fermi}-Large Area Telescope light curves of 11 blazars. The data were comprehensively investigated using the methods of nonlinear time series analysis: phase-space reconstruction, fractal dimension, maximal Lyapunov exponent (mLE). We tested several possible parameters affecting the outcomes, in particular the mLE, in order to verify the spuriousness of the outcomes. We found no signs of chaos in any of the analyzed blazars. Blazar variability is either truly stochastic in nature, or governed by high-dimensional chaos that can often resemble randomness.
\end{abstract}




\section{Introduction}

The \textit{Fermi}-Large Area Telescope \citep[LAT;][]{Atwo09} is a high energy $\gamma$-ray telescope, sensitive to photons in the energy range from 20 MeV to 300 GeV, which detected 5065 sources in the 100 MeV--100 GeV energy range \citep{Abdollahi}. More than 3130 sources were identified as blazars, a subclass of active galactic nuclei (AGNs), possessing a set of characteristic properties such as strong continuous radiation observed throughout the electromagnetic spectrum, flat-spectrum radio core, fast variability in any energy band, and a high degree of optical-to-radio polarization. In the unification scheme introduced by \cite{Urry95}, blazars are AGNs pointing their relativistic jets toward the Earth \citep[see e.g.][for a review]{Urry95,2012rjag.book.....B,Pado17}. Blazars are usually divided into two groups: BL Lacertae objects (BL Lacs) and flat spectrum radio quasars (FSRQs). This classification is historically based on the strength of the optical emission lines, i.e. FSRQs have broad emission lines with the equivalent width $>$5~\r{A}, while BL Lacs possess weak lines or no emission lines at all. Further classification is made taking into account position of a synchrotron peak, $\nu^s_{\mathrm{peak}}$ in the $\nu - \nu$F$_{\nu}$ plane, in the multiwavelength spectral energy distribution and different accretion regimes of AGNs. BL Lacs are commonly split into low-peaked, intermediate-peaked, and high-peaked (HBL) BL Lacs \citep{Abdo10}. An additional group of extreme HBLs, having $\nu_{\mathrm{peak}}^s\gtrsim 10^{17}\,{\rm Hz}$, is also considered \citep{Costamante01,Akiyama16}.

The search for chaos in AGNs has not been successful so far. One of the first attempts was done by \citet{lehto93}, who computed the correlation dimensions $d_C$ of the X-ray light curves (LCs) of eight AGNs, and reported evidence for $d_C<4.5$ for the Seyfert galaxy NGC~4051, suggesting that variability of this source might be chaotic in its nature.

\citet{provenzale94} investigated a long (800 days) optical LC of the quasar 3C 345 with the correlation dimension as well. While $d_C\approx 3.1$ was found, the authors demonstrated that this is a spurious detection owing to the long-memory property of the nonstationary signal driven by a power-law form of the power spectral density. They pointed at an intermittent stochastic process that produced outputs consistent with the observations. Therefore, the interpretation of any fractional correlation dimension of a phase-space trajectory reconstructed from a univariate time series needs to be backed up with additional evidence. The same technique was applied to microquasars \citep{misra04}, but the initially reported saturation of the correlation dimension was not found to be a signature of chaos, likely owing to the nonstationarity of the data again \citep{mannattil16}. Indeed, nonstationarity often leads to a spurious detection of chaos in a nonchaotic system \citep{tarnopolski15}, hence a proper transformation is required.

\citet{kidger96} performed microvariability analysis of the BL Lac 3C~66A in the optical and near infrared bands. They reported on a positive maximal Lyapunov exponent (mLE) and very low correlation dimensions, $d_C<2$. These are contradictory findings, since $1<d_C<2$ implies at most a two-dimensional phase space \citep{seymour13}, in which, according to the Poincar\'e-Bendixson theorem, chaos cannot occur \citep{lichtenberg92}. This can be most likely attributed to the very short LCs that were investigated. \citet{sadun96}, in turn, conducted a broad nonlinear time series analysis of the optical LCs of the famous OJ~287 double black hole (BH) system, and reported $2\lesssim d_C\lesssim 4$, with positive mLEs as well. The particular method for their calculation was not described explicitly, but it should be mentioned that the algorithm of \citet{wolf85}, frequently employed in the past, is biased towards detecting positive mLEs, especially for short data sets, since it does not test for exponential divergence, but assumes it explicitly ad hoc \citep{tarnopolski15}. A more rigorous, up-to-date analysis of OJ~287 is therefore appropriate.

Finally, most recently \citet{bachev15} analyzed a long (1.6 year), densely sampled (160\,000 points) optical Kepler LC of the BL Lac W2R 1926+42. They aimed to constrain the correlation dimension of the reconstructed phase-space trajectory, however the dimension did not saturate at any value smaller than the maximal tested embedding dimension $m=10$. Overall, a saturated or even fractional $d_C$ need not be due to the underlying chaotic dynamics, and hence a larger suite of nonlinear time series analysis techniques should be invoked, especially aiming at establishing the sign of the mLE, with a careful consideration of the stationarity of the analyzed data.

In this work we search for signatures of chaos in the $\gamma$-ray LCs of some of the brightest or otherwise famous blazars from \textit{Fermi}-LAT. We examine five BL Lacs (Mrk~501, Mrk~421, PKS~0716+714, PKS~2155-304, TXS~0506+056) and six FSRQs (PKS~1510-089, 3C~279, B2~1520+31, B2~1633+38, 3C~454.3, PKS~1830-211), i.e. the sample from \citet{tarnopolski20}. The methodology for studying chaotic behavior includes well-established methods of nonlinear time series analysis, such as reconstruction of the phase-space, correlation dimension, and mLE. We utilize the method of surrogates to ascertain the reliability of the results.

\section{Data}
\label{sect2}

To ensure stationarity, we investigated the logarithmized LCs in the 7-day binning in order to maximize the number of points \citep{tarnopolski20}, i.e. we seek for chaotic behavior in the process $l(t)$ underlying the observed variability $f(t)$. The two are connected via $f(t)=\exp[l(t)]$ since \citep{uttley05}:
\begin{enumerate}
\item the distribution of fluxes is lognormal,
\item the root mean square--flux relation is linear.
\end{enumerate}

\subsection{Fermi data}

We performed a spectral analysis of $\sim$11-year \textit{Fermi}-LAT data of 11 well known blazars, spanning between 54682 and 58592 MJD in an energy range of 100 MeV--300 GeV. We analyzed the data using a binned maximum likelihood approach\footnote{\url{https://fermi.gsfc.nasa.gov/ssc/data/analysis/scitools/binned\_likelihood\_tutorial.html}} in a region of interest (ROI) of $10\dg$ around the position of each blazar, with the latest 1.2.1 version of \textsc{Fermitools}, namely conda distribution of the Fermi ScienceTools\footnote{\url{https://github.com/fermi-lat/Fermitools-conda/wiki}}, and \textsc{fermipy} \citep{Wood17}. We used the reprocessed Pass 8 data and the $\text{P8R3}\_\text{SOURCE}\_\text{V2}$ instrument response functions. A zenith angle cut of $90^\circ$ is used together with the EVENT\_CLASS = 128 and the EVENT\_TYPE = 3, while the \emph{gtmktime} cuts DATA\_QUAL==1 $\&\&$ LAT\_CONFIG==1 were chosen. We defined the spatial bin size to be 0\fdg1, and the number of energy bins per decade of 8. The diffuse components\footnote{\url{https://fermi.gsfc.nasa.gov/ssc/data/access/lat/BackgroundModels.html}} were modeled with the Galactic diffuse emission model \texttt{gll\_iem\_v07.fits} and the isotropic diffuse model \texttt{iso\_P8R3\_SOURCE\_V2\_v01.txt}, including also all known point-like foreground/background sources in the ROI from the \textit{LAT 8-year Source Catalog} \citep[4FGL; ][]{Fermi19}. The LCs of each blazar were generated using 7-day time bins and selecting observations with the test statistic $TS>25$.

\subsection{Interpolation of missing points}

Data loss introduces a bias in the estimated frequency content of the signal, because the observed power spectrum is the result of the convolution of the true power spectrum with the spectral window function. Thus, recovering the entire duty cycle is necessary to identify the signatures of chaos in the LCs without biases.

Missing data points in the LCs were interpolated using the method of interpolation by autoregressive moving average algorithm \citep[MIARMA; ][]{PG15}, which is aimed to preserve the original frequency content of the signal. This algorithm makes use of a forward-backward predictor based on ARMA modeling. A local prediction is obtained for each interpolation allowing also that weakly nonstationary signals can be interpolated.

\section{Methods}
\label{sect3}
The whole analysis was performed for logarithmic LCs. We conducted the analysis of surrogates as well to make sure our results are not due to a chance occurrence. Every object was nonlinearly denoised (see Sect.~\ref{sect3.2}) before the phase-space reconstruction and the subsequent search for a positive mLE. The routines implemented in the {\sc TISEAN} 3.0.1 package\footnote{\url{https://www.pks.mpg.de/~tisean/}} \citep{TISEAN} were utilized throughout.

\subsection{Phase-space reconstruction}
\label{sect3.1}

The phase-space representation of a dynamical system is one of the key points in nonlinear data analysis. In theory, a dynamical system can be defined by a set of first-order ordinary differential equations that can be directly investigated to rigorously describe the structure of the phase space \citep{Kantz}. However, in case of real-world dynamical systems, the underlying equations are either too complex, or simply unknown. Observations of a physical process usually do not provide all possible state variables. Often just one observable is available, e.g. a series of flux values that form a LC. Such a univariate time series can still be utilized to reconstruct the phase space. 

A basic technique is to reconstruct the phase-space trajectory via Takens time delay embedding method \citep{Takens}. Having
a series of scalar measurements $x(t)$, evenly distributed at times $t$, one can form an $m$-dimensional location vector of delay coordinates, $S(t)$, using only the values of $x(t)$ according to 
\begin{equation}
\vec{S}(t) = [ x(t), x(t + \tau), x(t+2 \tau), ..., x(t + (m-1)\tau) ].
\label{eq1}
\end{equation}
The main difficulty while attempting the phase-space reconstruction lies in determining the values of the time delay $\tau$ and embedding dimension $m$. These can be obtained with the help of mutual information (MI, see Sect.~\ref{sect3.3}) and the fraction of false nearest neighbors (FNN, see Sect.~\ref{sect3.4}). To uncover the structure buried in observational fluctuations, noise reduction techniques are also employed.

\subsection{Nonlinear noise reduction}
\label{sect3.2}

Generally, noise reduction methods for nonlinear chaotic time series work iteratively. In each iteration the noise is repressed by requiring locally linear relations among the delay coordinates, i.e. by moving the delay vectors toward some smooth manifold. We performed noise reduction with the algorithm designed by \citet{Grassberger}, implemented as the \texttt{ghkss} routine in the {\sc TISEAN} package. The concept is as follows: the dynamical system forms a $q$-dimensional manifold $M_1$ containing the phase-space trajectory. According to the Takens' embedding theorem there exists a one-to-one image of the path in the embedding space, if $m$ is sufficiently high. Thus, if the measured time series was not corrupted with noise, all the embedding vectors $\vec{v}_n$ would lie inside another manifold $M_2$ in the embedding space. However, due to the noise this condition is no longer fulfilled. The idea of the locally projective noise reduction scheme is that for each $\vec{v}_n$ there exists a correction $\Theta_n$, with $|| \Theta_n ||$ small, in such a way that $\vec{v}_n - \Theta_n \in M_2$ and that $\Theta_n$ is orthogonal on $M_2$. Of course a projection to the manifold can only be a reasonable concept if the vectors are embedded in spaces which are higher-dimensional than the manifold $M_2$. Thus we have to over-embed in $m$-dimensional spaces with $m>q$. 

With the metric tensor $G$ defined as
\begin{equation}
G_{ij} = 
\begin{cases}
 1 & i = j,\, i > 1,\, j < m \\
 0 & {\rm otherwise}
\end{cases},
\label{}
\end{equation}
where $m$ is the dimension of the ''over-embedded'' delay vectors, the minimization problem $\sum\limits_i (\Theta_i G^{-1} \Theta_i) = min$ is to be solved, including the following constraints: 
\begin{enumerate}
\item $a^{i}_{n}(\vec{v}_n - \Theta_n) + b^{i}_{n} = 0$ (for $i = q+1, ..., m$);
\item $a^{i}_{n} G a^{i}_{n} = \delta_{ij}$.
\end{enumerate}
where the $a^{i}_{n}$ are the normal vectors of $M_2$ at the point $\vec{v}_n-\Theta_n$; $b^i_n$ could be found solving a minimization problem \citep{Grassberger}, where $b^i_n = - a^i_n \cdot \xi^i$ and $\xi$ is given as a linear combination $\xi^i_n = \sum_{v_k \in U_n} \omega_k v_{k+n} $. The neighborhood for each point $\vec{v_n}$ is $U_n$, and $\omega_k$ is a weight factor with $\omega_k > 0 $ and $ \sum_k \omega_k = 1$.

\subsection{Mutual Information (MI)}
\label{sect3.3}

The most reasonable delay is chosen as the first local minimum of the MI. The $\tau$ time delayed MI is defined as
\begin{equation}
I_{i,j}(\tau) = -\sum_{i,j=1}^{n} {P_{ij}(\tau)\ln{\frac{P_{ij}(\tau)}{P_{i}P_{j}}}},
\label{eq2}
\end{equation}
 where $P_{ij}(\tau)$ is the joint probability that an observation falls in the $i$-th interval and the observation time $\tau$ falls in the $j$-th interval, $P_{i}$ and $P_{j}$ are the marginal probabilities \citep{fraser86}. In other words, it gives the amount of information one can obtain about $x_{t + \tau}$ given $x_t$. The absolute difference between $x_{\rm max}$ and $x_{\rm min}$ of the data is binned into $n$ bins, and for each bin the MI as a function of $\tau$ is constructed from the probabilities that the variable lies in the $i$-th and $j$-th bins and the $P_{ij}(\tau)$ that $x_t$ and $x_{t + \tau}$ are in the $i$-th and $j$-th bins, respectively \citep{tarnopolski15}.
 
Additionally, we used also the autocorrelation function (ACF), with the criterion to choose as the delay $\tau$ the first lag at which the ACF drops below $1/{\rm e}$, but the obtained delays did not always match with those from the MI. Therefore, all delays inferred from MI, ACF, and values in between were checked in subsequent steps of the analysis. Some parameters could not give a clear interpretation as per the chaotic behavior, though. Both MI and ACF were implemented within the \texttt{mutual} routine in the {\sc TISEAN} package.

\subsection{False Nearest Neighbors (FNN)}
\label{sect3.4}

The FNN method is a way of determining the minimal sufficient embedding dimension $m$. This means that in an $m_0$-dimensional delay space the reconstructed trajectory is a topological one-to-one image of the trajectory in the original phase space. If one selects a point on it, then its neighbors are mapped onto neighbors in the delay space. Thus the neighborhoods of points are mapped onto neighborhoods, too. However, the shape and diameter of the neighborhoods vary depending on the LEs. But if one embeds in an $m$-dimensional space with $m<m_0$, points are projected onto neighborhoods of other points to which they would not belong in higher dimensions (aka false neighbors), because the topological structure is no longer retained. The FNN algorithm looks for nearest neighbor $\vec{k}_j$, for each point $\vec{k}_i$ in an $m$-dimensional space, and calculates the distance $||\vec{k}_i - \vec{k}_j||$. It then iterates over both points and computes 
\begin{equation}
R_{i} = \frac{||\vec{k}_{i+1} - \vec{k}_{j+1}||}{||\vec{k}_i - \vec{k}_j||}.
\label{}
\end{equation}
Thereby, a false neighbor is any neighbor for which $R_{i} > R_{\rm tol}$, where $R_{\rm tol}$ is some threshold. This algorithm was firstly proposed by \citet{kennel92}, and next improved by \citet{Ka}. The FNN algorithm is widely used for detecting chaotic behavior in data sets obtained from astrophysical observations \citep{Hanslmeier}, to experimental measurements connected with electronics \citep{U}. We utilized the \texttt{false\_nearest} routine from the {\sc TISEAN} package.

\subsection{Lyapunov exponent}
\label{sect3.5}
The LE is one of the main characteristics in the analysis of chaotic dynamical system. The LE characterizes the rate of separation of infinitesimally close trajectories \textbf{Z}(t) and $ \textbf{Z}_0(t) $ in the phase space \citep{Cecconi}. It describes the evolution of the separation $\delta \textbf{Z}(t)= \textbf{Z}(t) - \textbf{Z}_0(t)$ via
\begin{equation}
| \delta \textbf{Z}(t)| \approx e^{\lambda t }| \delta \textbf{Z}_0(t)|, 
\end{equation}
where $\delta \textbf{Z}_0(t) = \textbf{Z}(0) - \textbf{Z}_0(0)$. The mLE is a measure of predictability for a given solution to a dynamical system, and is formally determined as: 
\begin{equation}
\lambda_{\rm max} = \lim_{t\to\infty}\lim_{\delta\textbf{Z}_0\to0} \frac{1}{t} \ln \frac{| \delta \textbf{Z}(t)|}{| \delta \textbf{Z}_0(t)|}
\end{equation}
A positive mLE usually indicates that the system is chaotic, i.e. exhibits sensitive dependence on initial conditions, manifesting itself through exponential divergence.

For the estimation of the mLE of a given univariate time series set we use Kantz method \citep{TISEAN} in our analysis, implemented as the \texttt{lyap\_k} routine in the {\sc TISEAN} package. The algorithm takes points in the neighborhood of some point $x_i$. Next, it computes the average distance of all acquired trajectories to the reference, $i$-th one, as a dependence of the relative time $n$. The average $S(n)$ of the logarithms of these distances (so-called stretching factors) is plotted as a function of $n$. In the case of chaos, three regions should be distinct: a steep increase for small $n$, a linear part and a plateau \citep{seymour13}. The slope of the linear increase is the mLE; its inverse is the Lyapunov time.

\subsection{Correlation dimension}
\label{sect::dC}

A fractal dimension \citep{mandelbrot83} is often measured with the correlation dimension, $d_C$ \citep{grassberger83}, which takes into account the local densities of the points in the examined dataset. For usual 1D, 2D or 3D cases the $d_C$ is equal to 1, 2 and 3, respectively. Typically, a fractional correlation dimension is obtained for fractals.\footnote{Although some fractals can exhibit integer fractal dimensions, just different from the embedding dimension; e.g. the boundary of the Mandelbrot set has a dimension of exactly 2 \citep{shishikura98}.}

The correlation dimension is defined as
\begin{equation}
d_C=\lim_{R\rightarrow 0}\frac{\ln C(R)}{\ln R},
\label{eqCorr}
\end{equation}
with the estimate for the correlation function $C(R)$ being
\begin{equation}
C(R)\propto\sum_{i=1}^N\sum_{j=i+1}^N \Theta\left(R-||x_i-x_j||\right),
\label{eqB}
\end{equation}
where the Heaviside step function $\Theta$ adds to $C(R)$ only points $x_i$ in a distance smaller than $R$ from $x_j$ and vice versa. The total number of points in the reconstructed phase-space trajectory is denoted by $N$, and the usual Euclidean distance is employed. The limit in Eq.~(\ref{eqCorr}) is attained by fitting a straight line to the linear part of the obtained $\log C(R)$ vs. $\log R$ dependency. The dimension $d_C$ is estimated as the slope of this linear regression.

\citet{eckmann92} argued that for a time series of length $N$, the maximal meaningful value of $d_C$ is necessarily less than $2\log N$ \citep[see also][]{ruelle90}. For the LCs examined herein, we have $N\gtrsim 500$, hence $d_C\lesssim 5$. We therefore search for low-dimensional chaos, i.e. with $m\sim 3-5$.

\subsection{Surrogate data}
\label{sect3.6}

The method of surrogates is the most commonly employed one to provide a reliable statistical evaluation in order to ensure that the observed results are not obtained by chance, but are a true characteristic of the system. Surrogates can be created as a data set that is generated from a model fitted to the observed (original) data, or directly from the original data (by some suitable transformation of it). Testing for the underlying nonlinearity with surrogates requires an appropriate null hypothesis: the data are linearly correlated in the temporal domain, but are random otherwise. In our employed approach, surrogates are generated from the original data while destroying any nonlinear structure by randomizing the phases of the Fourier transform \citep{Theiler,Oprisan}.
 
We use the routine \texttt{surrogates} from the TISEAN package, that generates multivariate surrogate data (i.e., implements the iterative Fourier scheme). The idea behind this routine is to create a whole ensemble of different realizations of a null hypothesis, and to apply statistical tests to reject the null for a given data set. The algorithm creates surrogates with the same Fourier amplitudes and the same distribution of values as in the original data set \citep{Kantz}.
If the chaotic signature is present in the original data, but not in the surrogates, one can ascertain that the detection of chaotic behavior is a real phenomenon.

\section{Results}
\label{sect4}

The 11 blazars in our sample were examined according to the methodology outlined in Sect.~\ref{sect3}. We cannot claim the presence of chaos in any of the analyzed objects. In the following we illustrate the analysis with an example of one blazar, i.e. 3C~279, leading to the conclusion of the lack of chaotic behavior in this source. Similar results were obtained for the remaining 10 blazars.

\subsection{Embedding dimension $m$}
\label{sect4.1}

\begin{figure}
\centering
\includegraphics[width=0.8\linewidth]{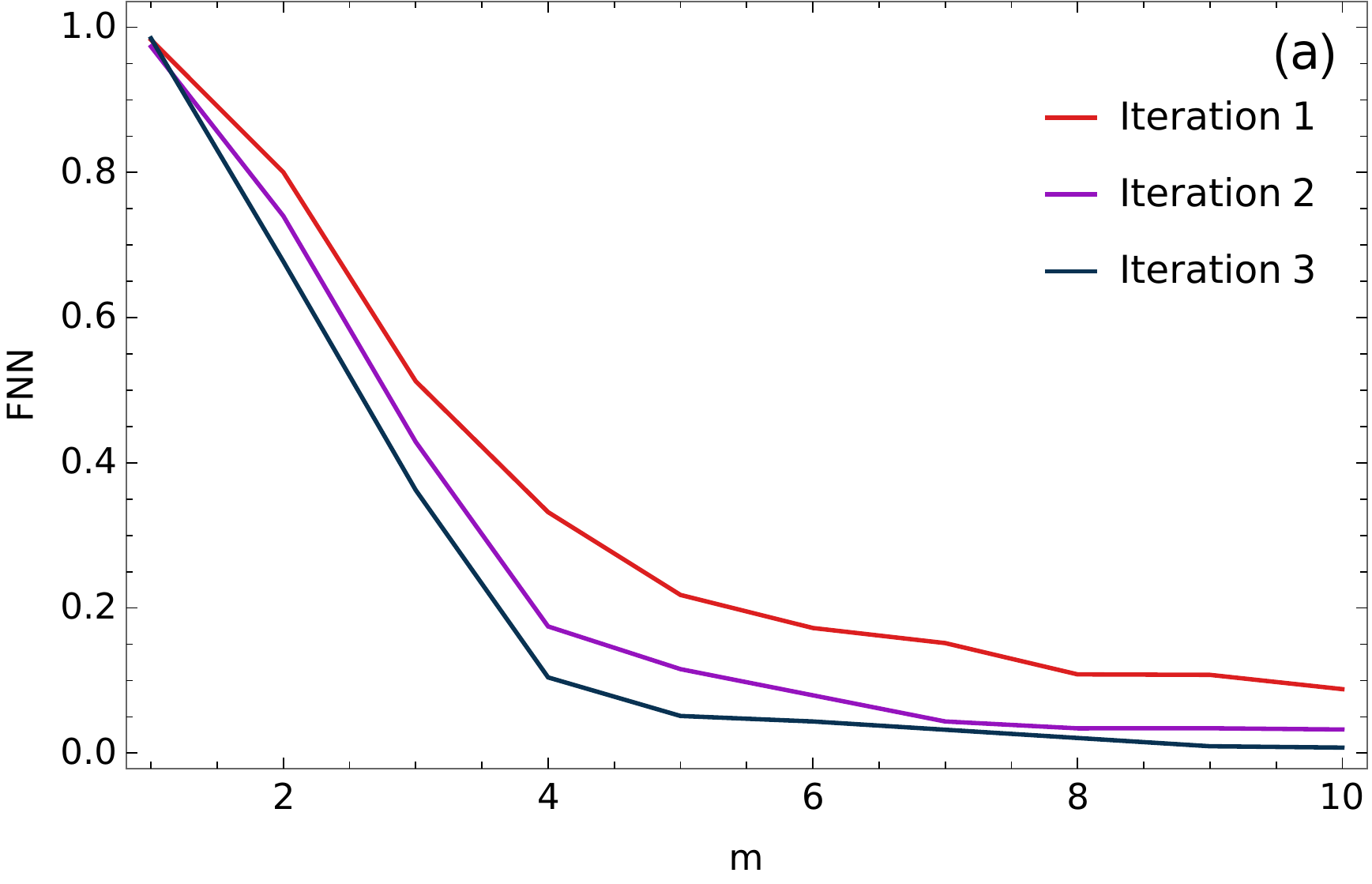}
\includegraphics[width=0.8\linewidth]{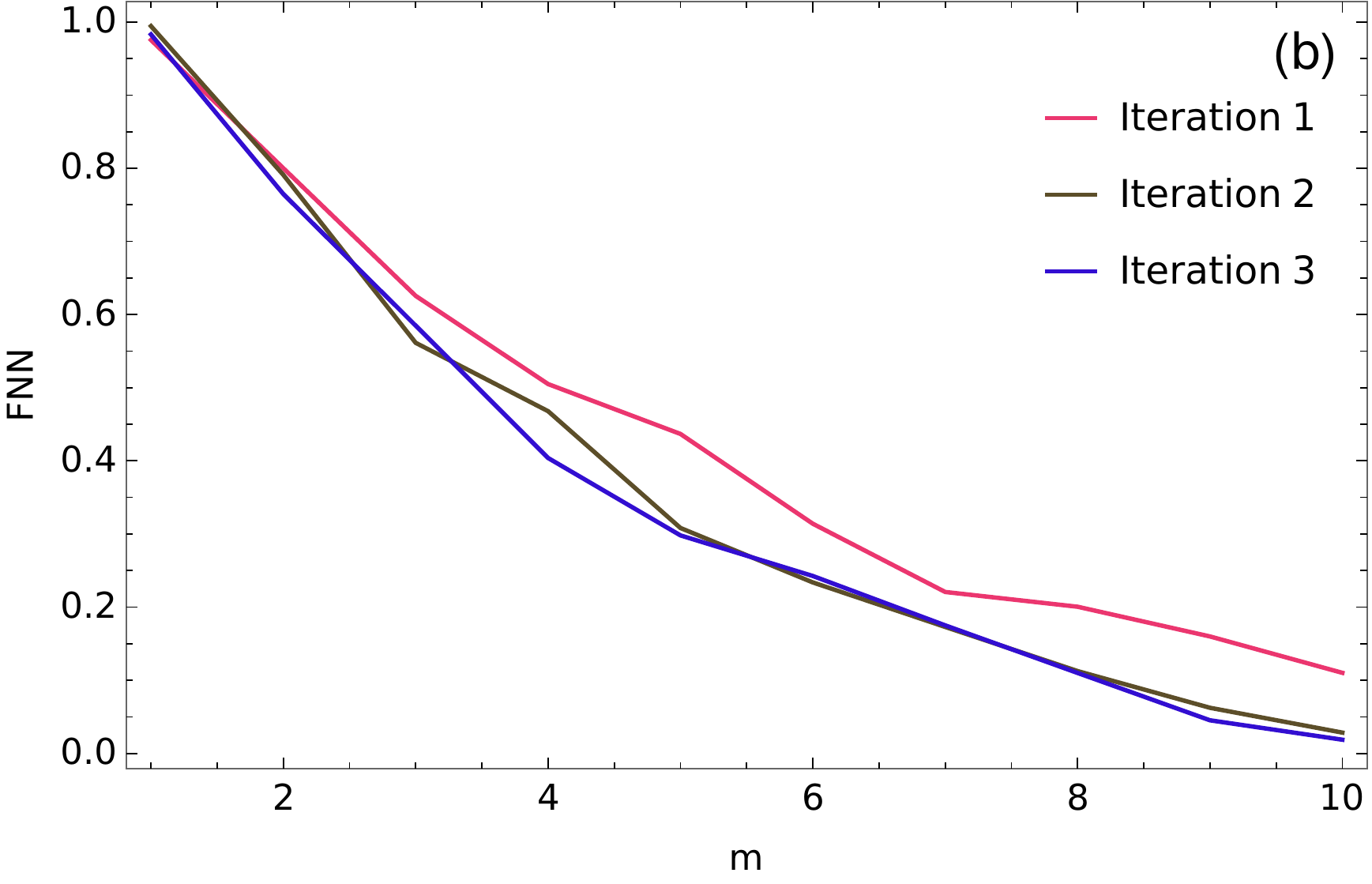}
\includegraphics[width=0.8\linewidth]{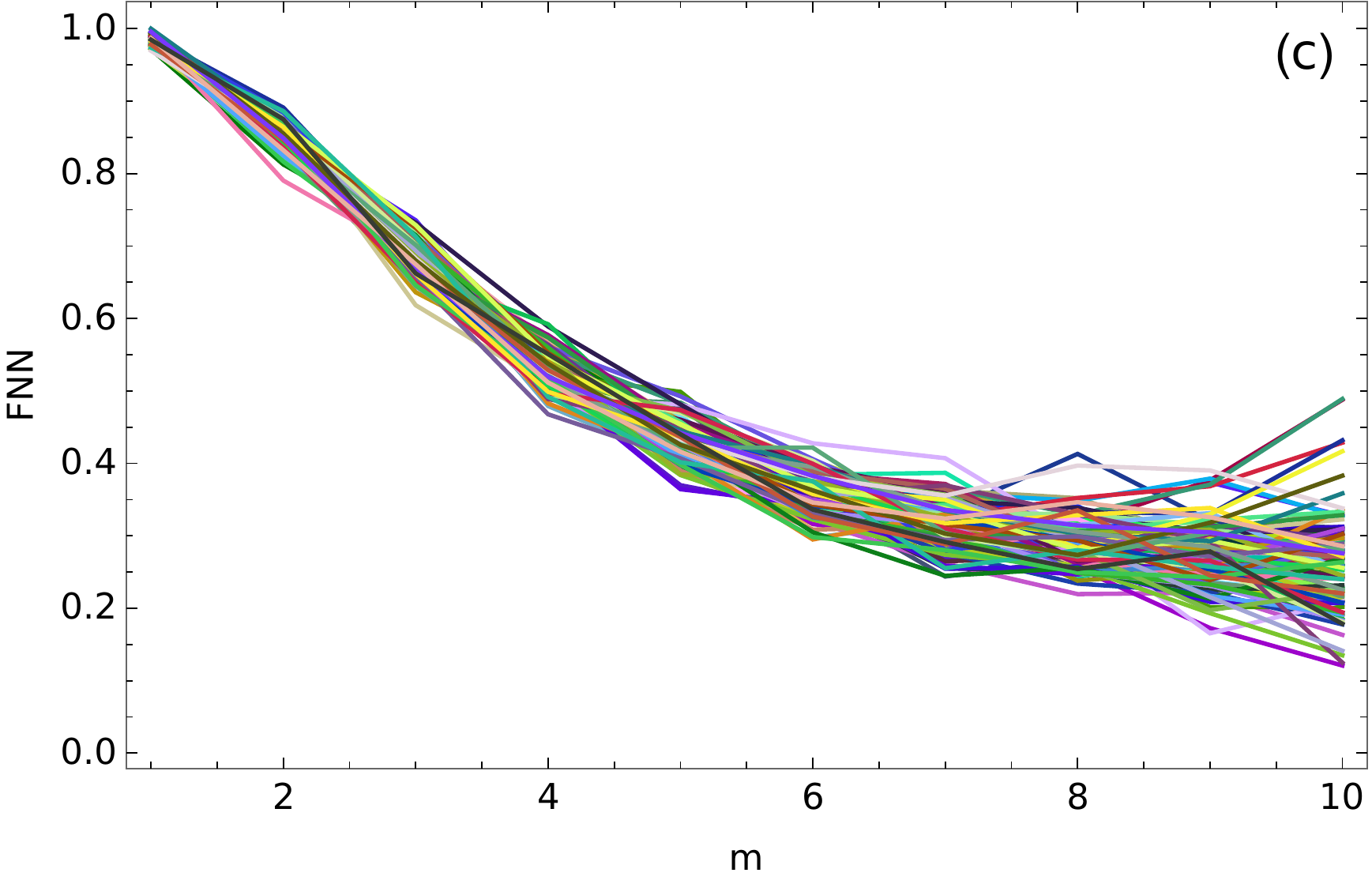}

\caption{The FNN plot for the logarithmic LC of 3C~279. (a) The knee at $m\simeq 4-5$ indicates the most appropriate embedding dimension; the value $\tau = 3$ was used. (b) One can not see such a sharp bending when was set to the delay value $\tau = 9$. (c) The lack of a sharp bending is also evident in case of the surrogates.} 
\label{fig:1}
\end{figure}

The FNN algorithm was employed to infer the proper embedding dimension $m$. The FNN fraction for different $m$ is displayed in Fig.~\ref{fig:1}. A clear bending (a knee) is seen at $m\simeq 4-5$ on the FNN plot (Fig.~\ref{fig:1}(a)). The three curves represent one, two, and three iterations of the denoising procedure of the original LC with the value $\tau = 3$. However, there is no clear bend on the FNN plot (b), which was obtained with the delay value $\tau = 9$. In order to ascertain that this result is not a chance occurrence, 100 surrogates were generated for every data set and their FNN fractions were computed. A representative subset of such surrogates and their mean value is displayed in Fig.~\ref{fig:1}(c). The FNN fractions remain high for all $m$ tested, and overall no clear bend is visible. 

\subsection{Time delay $\tau$}
\label{sect4.2}

\begin{figure}
\centering
\includegraphics[width=0.9\linewidth]{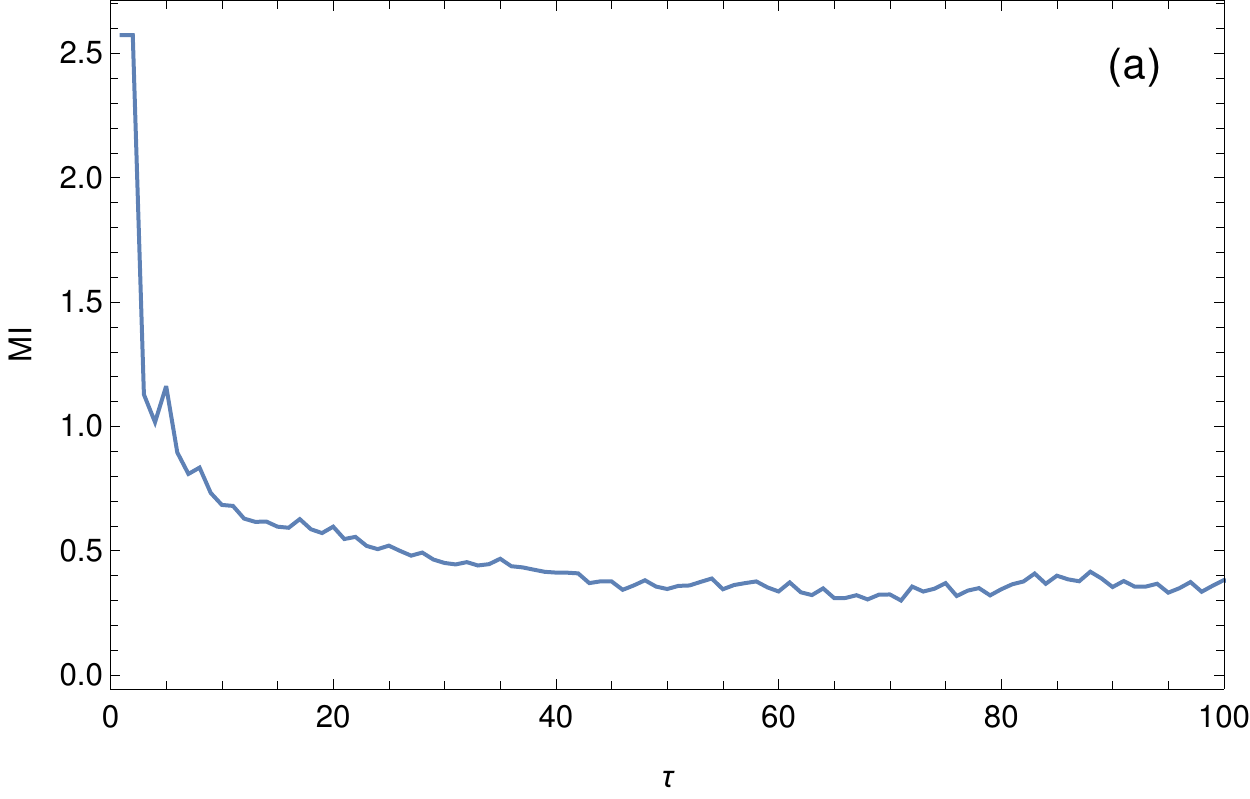}
\includegraphics[width=0.9\linewidth]{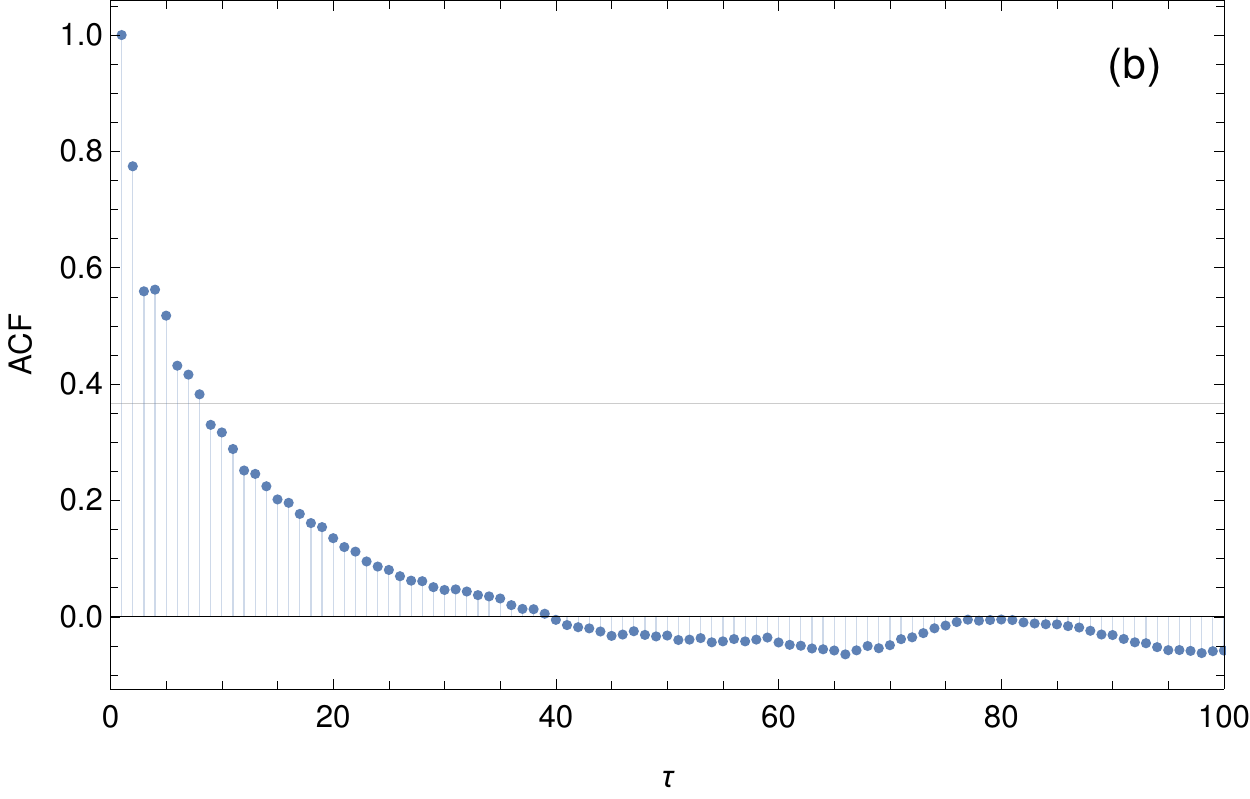}
\caption{Estimation of the delay $\tau$. (a) The MI has its first local minimum at $\tau=3$. 
(b) The ACF drops below $1/{\rm e}$ at $\tau=8$.}
\label{fig:2}
\end{figure}

After testing different values of $\tau$ for the phase-space reconstruction and the LEs, we came to the conclusion that various $\tau$ values lead to dramatically different results. MI yielded in general $\tau<15$. The same range of $\tau$ was implied from the ACF, although often the particular values were inconsistent. Fig.~\ref{fig:2} shows illustrative plots of MI and ACF. By investigating the phase-space reconstructions and the resulting mLEs, we settled using the values $\tau = 3$ and $\tau = 8$ as representative.

\subsection{Phase-space reconstruction}
\label{sect4.3}

\begin{figure}
\centering
\includegraphics[width=0.8\linewidth]{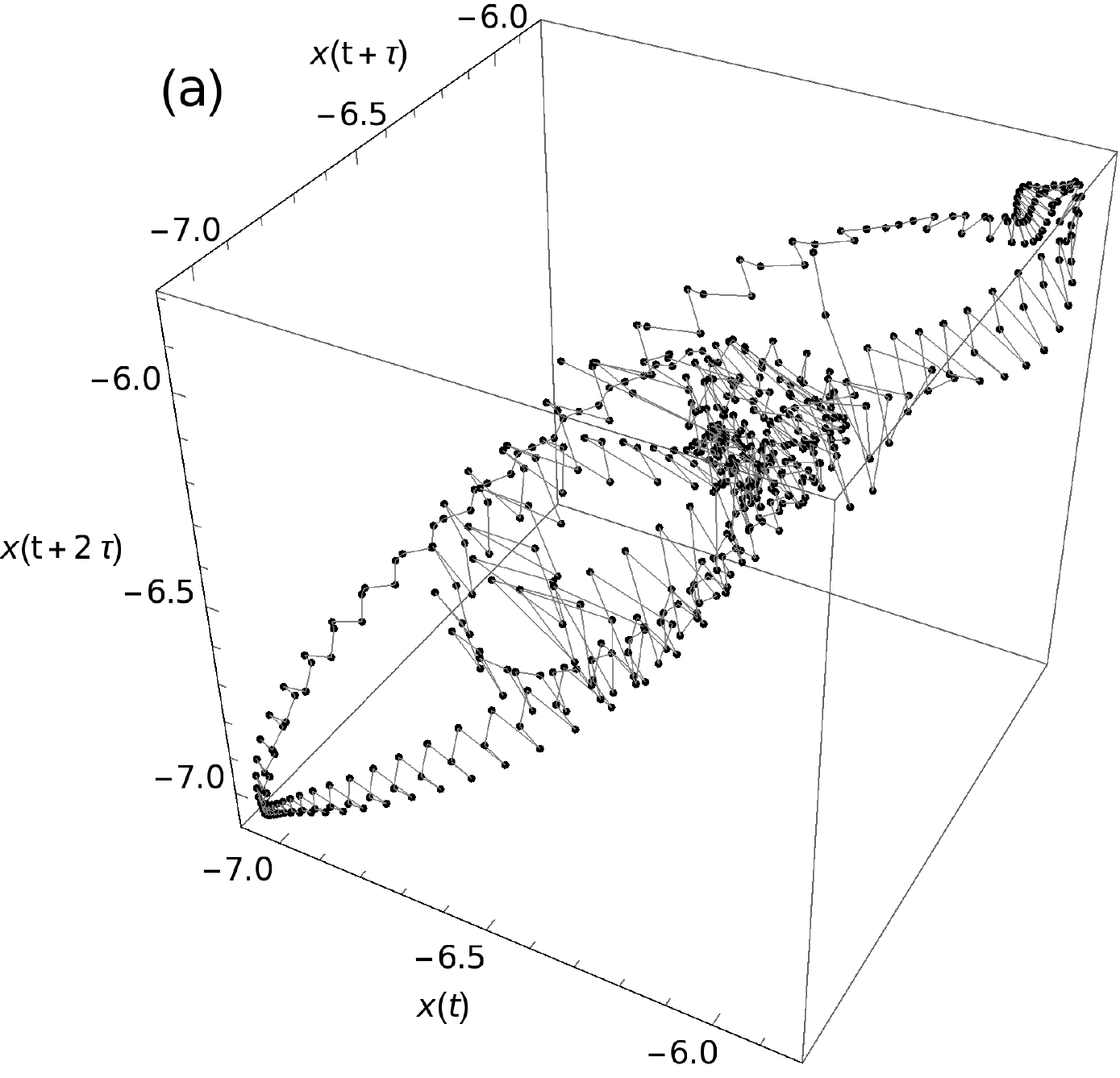}
\includegraphics[width=0.8\linewidth]{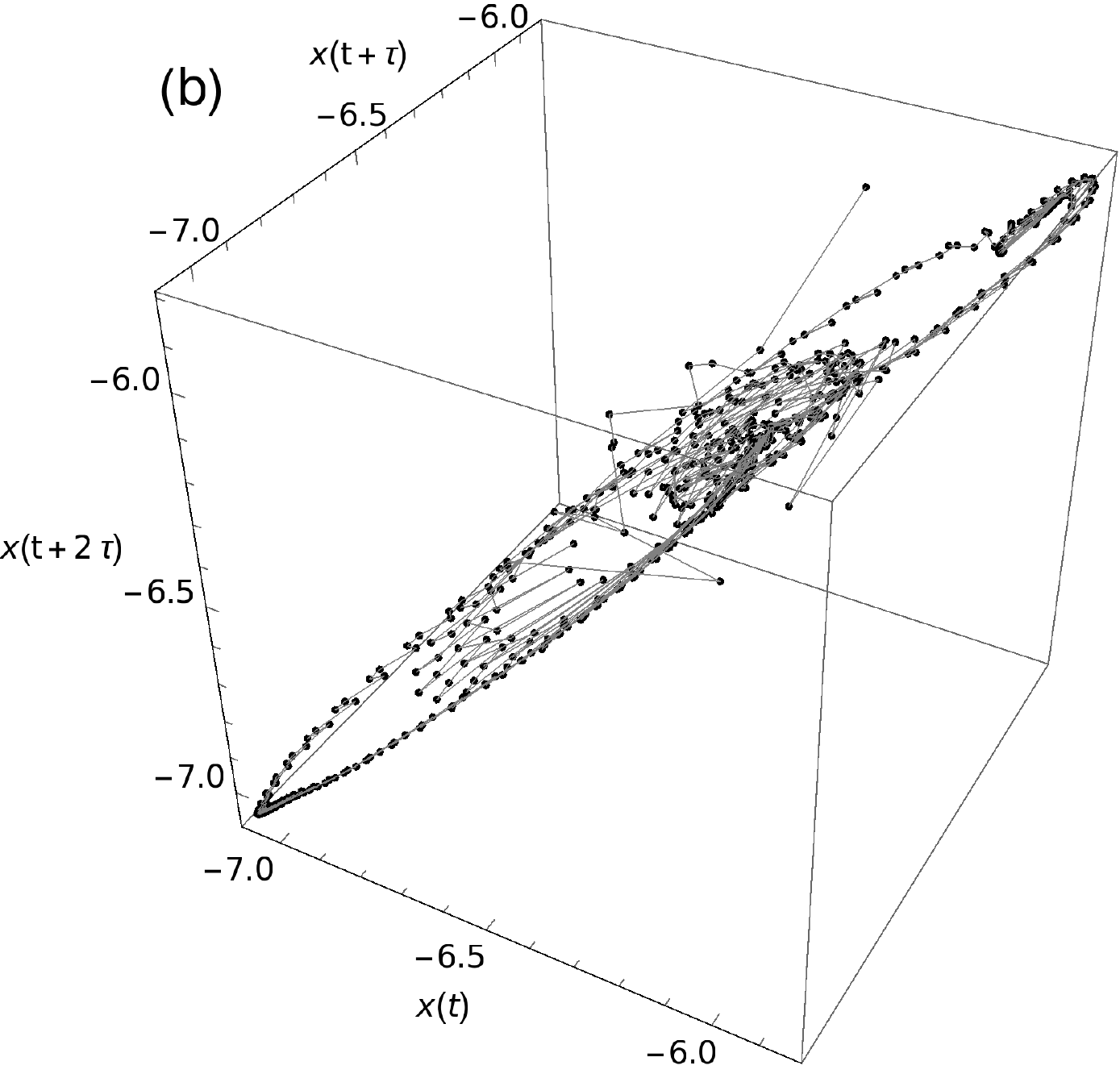}
\includegraphics[width=0.8\linewidth]{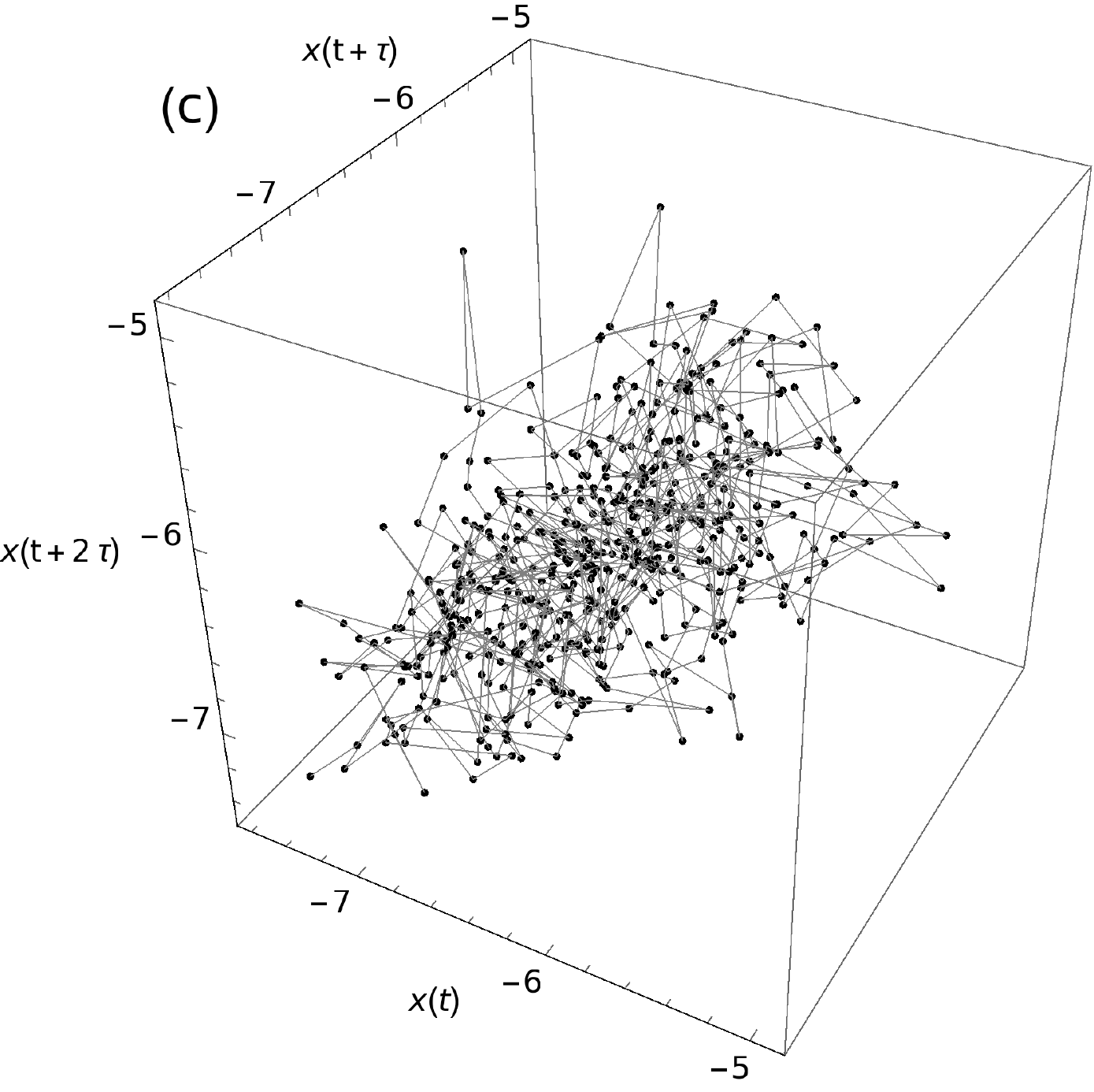}

\caption{A 3-dimensional phase-space reconstruction of 3C~279. These are projections of the underlying 5-dimensional trajectories. (a) The delay $\tau = 8$ was used. 
(b) In this case $\tau = 3$. The topology is still similar.
(c) A reconstruction of surrogates. Any structure was destroyed. }
\label{fig:3}
\end{figure}

With the obtained values of $m$ and $\tau$, one can in principle produce a phase-space reconstruction of the trajectory according to Eq.~(\ref{eq1}). However, for obvious reasons, illustrating graphically the resulting 4- or 5-dimensional trajectory is impossible. For display purposes only, a representation with $\tau = 3$ and $\tau =8$ in a 3-dimensional space is displayed in Fig.~\ref{fig:3}, together with a typical exemplary reconstruction of one of the surrogates.

\subsection{Maximal Lyapunov exponent}
\label{sect4.4}

\begin{figure}
\centering
\includegraphics[width=0.9\linewidth]{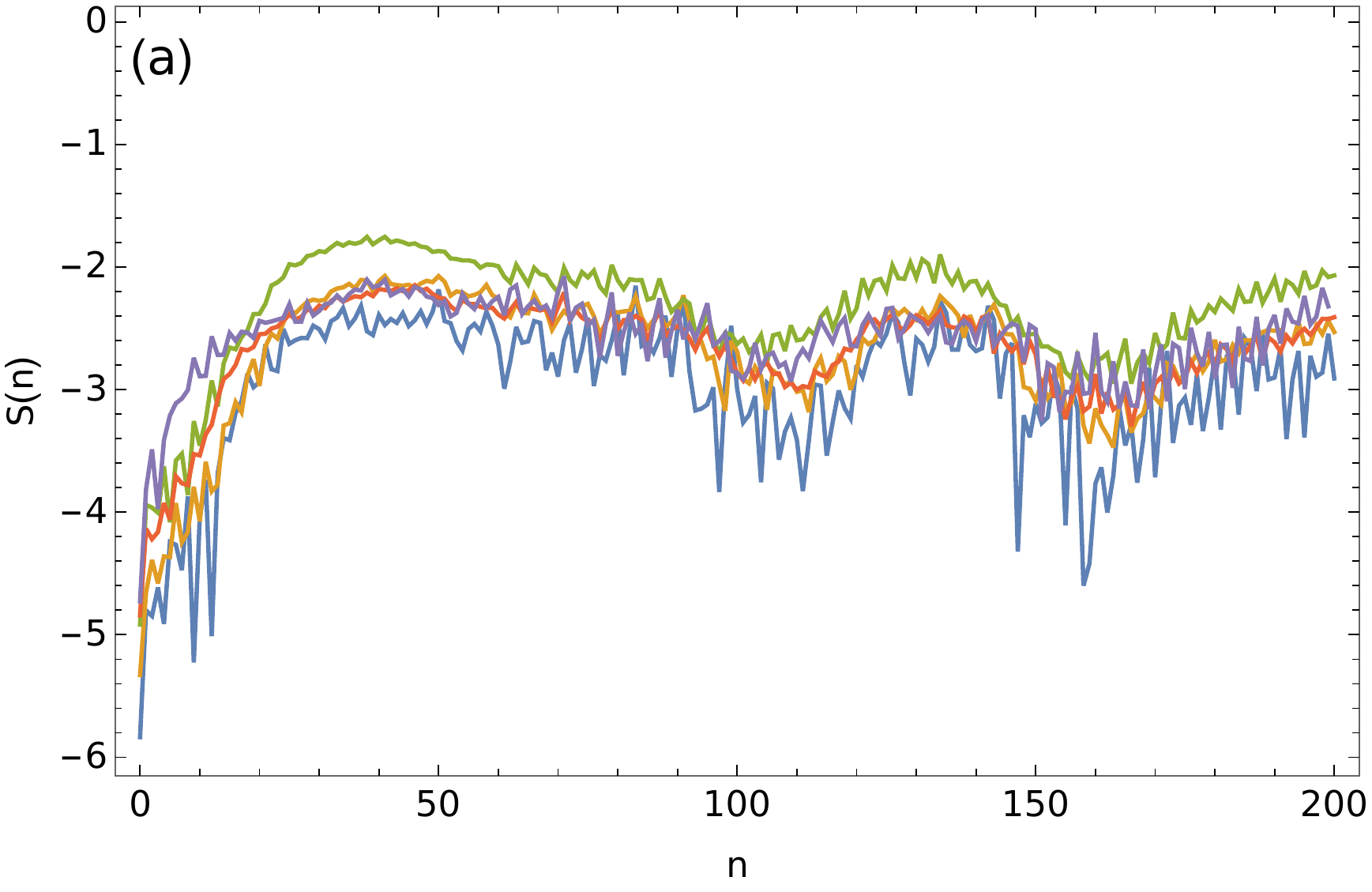}
\includegraphics[width=0.9\linewidth]{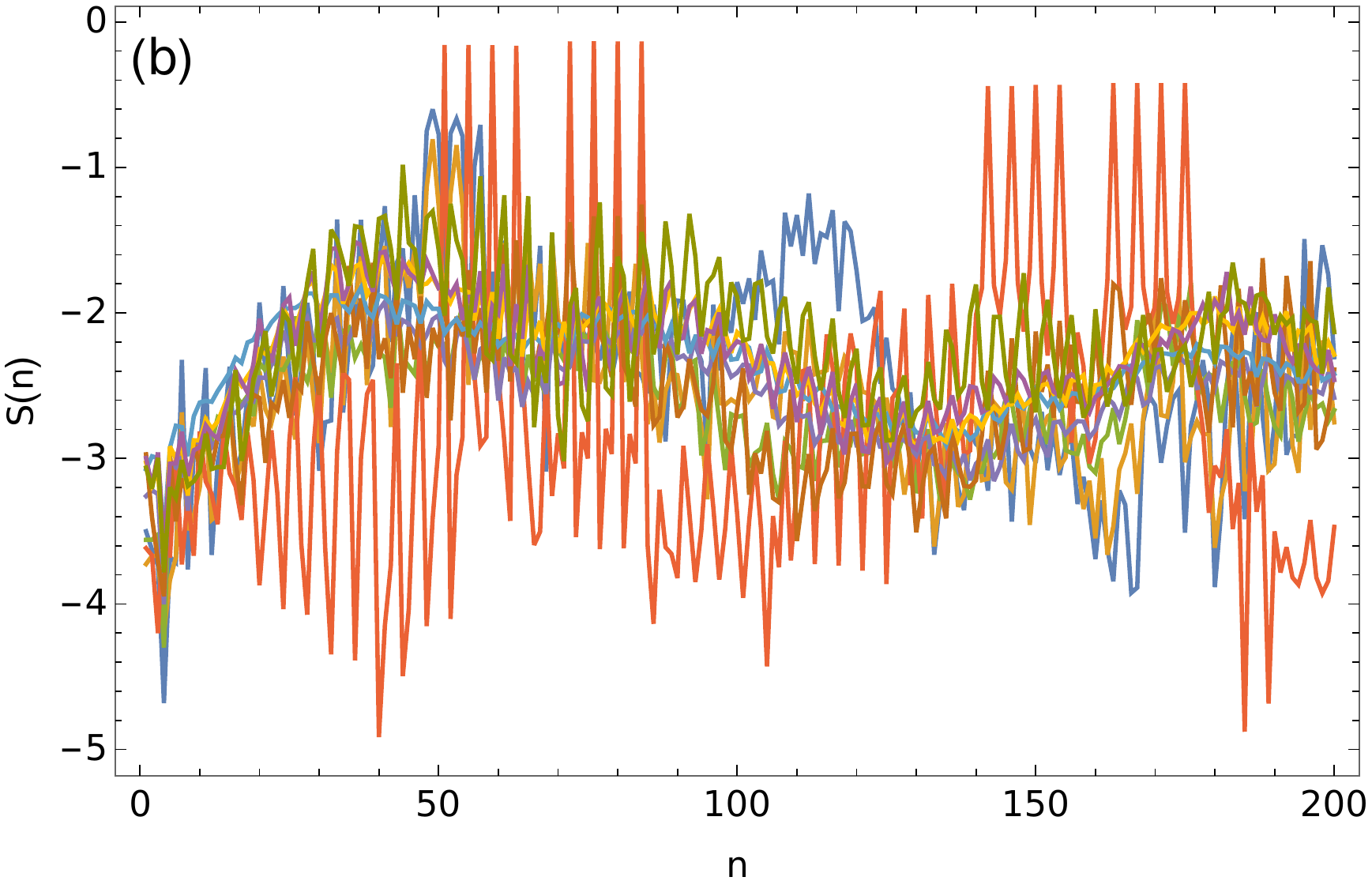}
\caption{The stretching factors $S(n)$ of (a) 3C~279 and (b) a representative surrogate. In both cases there is no unambiguous linear increase. Different colors correspond to different embedding dimensions $m$. }
\label{fig:4}
\end{figure}

Utilizing the obtained values of $m$ and $\tau$, we eventually attempted to constrain the mLE. In Fig.~\ref{fig:4}, the stretching factors $S(n)$ are depicted for the logarithmic LC itself, as well as for a representative example of a surrogate. As mentioned in Sect.~\ref{sect3.5}, in case of chaos three regions should be clearly visible: a sharp increase for very small $n$, followed by a linear section, and finally a plateau. None of these parts are present in Fig.~\ref{fig:4}(a), also no such features are present in any of the surrogates (cf. Fig.~\ref{fig:4}(b)).  Such results were arrived at for all 11 blazars in our sample.

\subsection{Correlation dimension}
\label{sect4.5}

\begin{figure}
\centering
\includegraphics[width=0.7\linewidth]{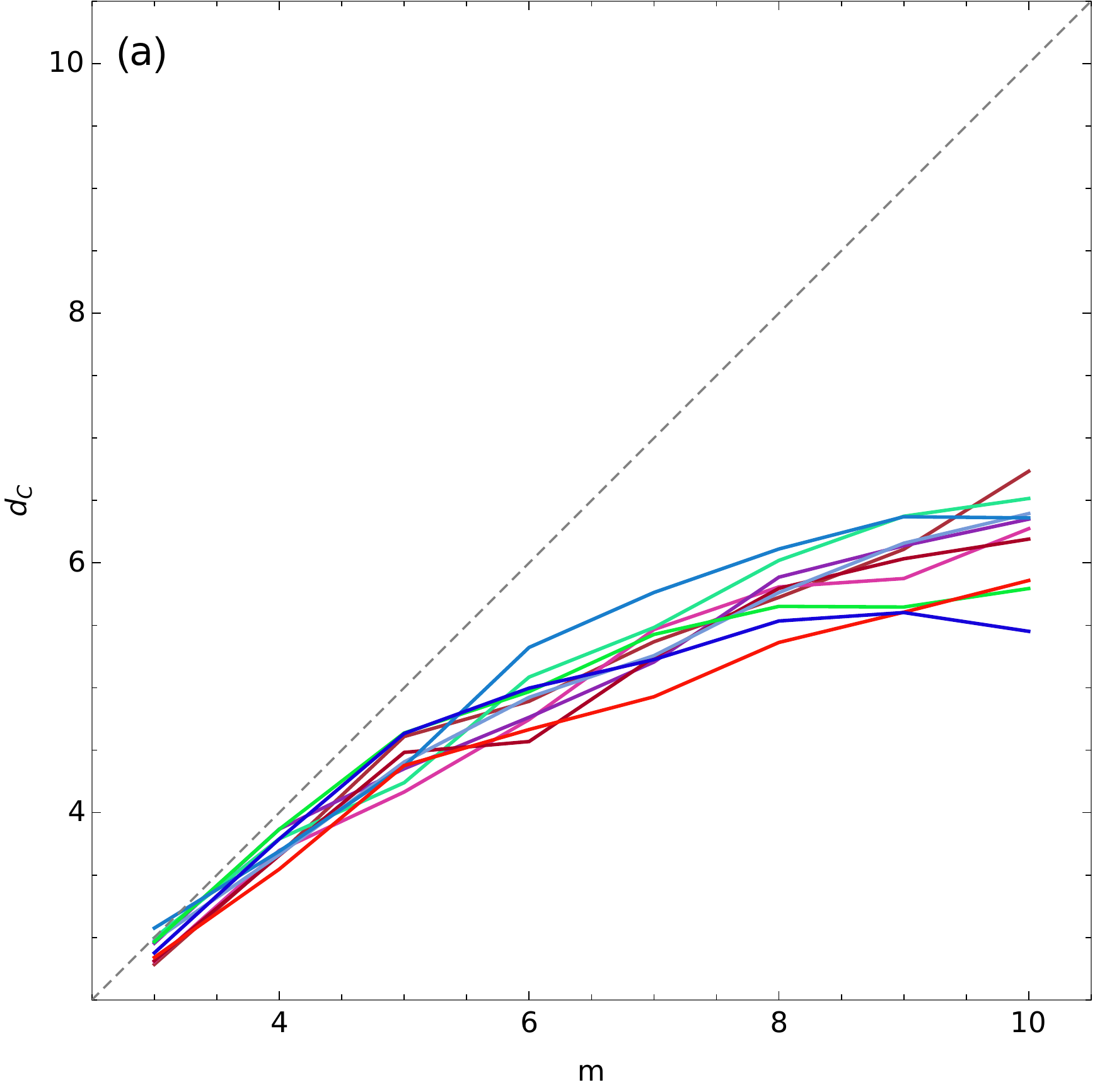}
\includegraphics[width=0.7\linewidth]{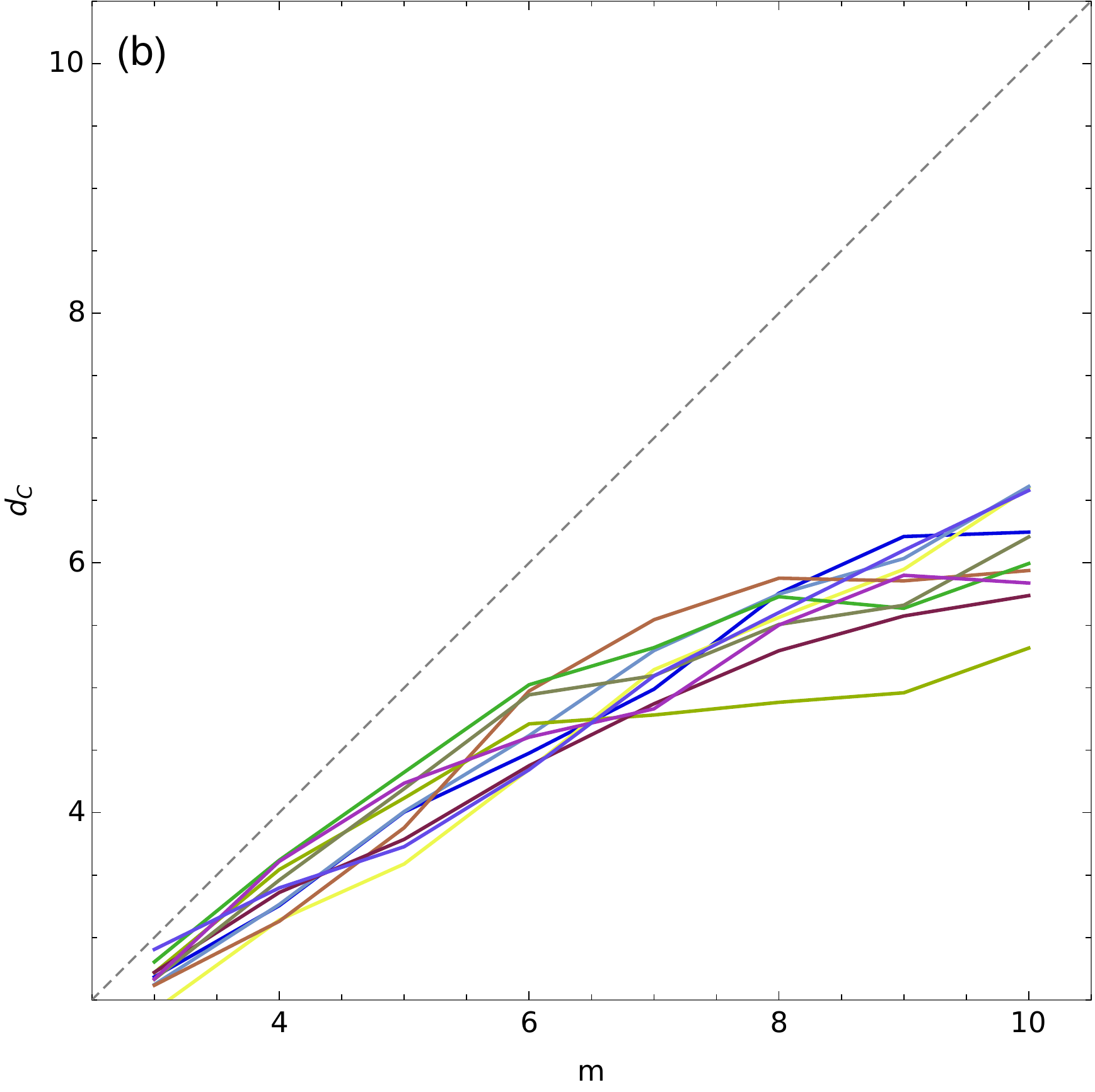}
\caption{ Correlation dimension $d_C$ for 3C~279 of (a) the logarithmic LC, and (b) of one of the surrogates. In both plots there is no clear plateau. }
\label{fig:5}
\end{figure}

We constructed a plot of $d_C$ as a function of $m$, which is presented in Fig.~\ref{fig:5}, as means of comparing with other works that utilized this method. In case of a chaotic system a linear increase followed by a plateau should be seen. The analysis of 3C~279 (Fig.~\ref{fig:5} (a)) did not provide evidence of chaotic behavior of the system. The plot of the surrogate data in Fig.~\ref{fig:5} (b) does not exhibit a plateau part as well. This observation applies to all 11 blazars considered herein.

\section{Discussion}

Finding low-dimensional chaos in a phenomenon with not well constrained physics is of great importance, since it provides information about the complexity of the underlying laws governing its occurrence \citep{seymour13,bachev15}. This in particular refers to blazar LCs, in which no unambiguous signs of chaos have been detected. Indeed, the analyses presented herein also did not give the slightest hints allowing to suspect the presence of chaos in any of the 11 objects examined. We displayed here the results corresponding to 3C~279; the other 10 blazars yielded very similar outcomes.

In principal, the behavior of a dynamical system can be described by a set of first-order ordinary differential equations. Such system can be directly investigated to uncover the structure of the phase space and to characterize its dynamical properties. Attractors, their fractal dimensions, Lyapunov exponents, etc. can be easily estimated, and their properties can be studied analytically, semi-analytically, and numerically. However, the underlying equations in most cases of real-world dynamical systems, such as blazar LCs, are unknown. Hence the detection of chaos, or lack of thereof, is a notoriously difficult task, especially in cases when the time series are relatively short and contaminated by observational noise. Noisy data can hinder the detection of chaos; moreover, high-dimensional chaos can be disguised as randomness. \citet{bachev15} argued that in the single-zone model there are only a handful of parameters that control the emission, which is governed by the Fokker-Planck (continuity) equation. However, if the radiation mechanism, and subsequently the variability of blazars, can indeed be accurately described by the (partial differential) continuity equation with some appropriate injection term \citep{stawarz08,finke14,finke15,chen16}, then the dynamical system is considered to be infinite-dimensional. While chaos can be present in infinite-dimensional systems (e.g. delayed systems; \citealt{Wernecke19}), its detection in astronomical LCs need not be unambiguously identifiable with the standard tools, most commonly applied with a discovery of low-dimensional chaos in mind---especially when the details of the fundamental dynamical processes remain unknown, and the time series is not extremely long.

On the other hand, if the radiation is influenced by turbulence in the jets, e.g. by chaotic magnetic flows caused by plasma instabilities, there might be instances in which the behavior of the system settles on some low-dimensional attractor. Therefore, further search for chaos in high-quality (multiwavelength) data gathered by next-generation space instruments, like the James Webb Space Telescope \citep{jwst}, can be expected to give a more definite answer. A uniform, rigorous analysis, in the spirit presented herein, of the already abundantly available optical data from the Kepler space telescope \citep{smith18} is also called for.

\section{Conclusions}

The aim of this paper was to search for evidence of low-dimensional chaos in $\gamma$-ray LCs of 11 blazars, i.e. five BL Lacs and six FSRQs. Data from \textit{Fermi}-LAT (10-year-long LCs with a 7-day binning) were investigated using the phase-space reconstruction via embedding dimension $m$ and time delay $\tau$, with the goal of eventually constraining the mLE (if positive) and correlation dimension $d_C$.

All analyses implied no signs of chaos for all 11 blazars. Therefore, the underlying physical processes that give rise to the observed variability are either truly stochastic \citep{tavecchio20}, or are governed by high-dimensional (possibly infinite-dimensional as well) chaos that can resemble randomness.

\section*{Acknowledgements}

O.O. thanks the Astronomical Observatory of the Jagiellonian University for the summer internship during which this research began.
M.T. acknowledges support by the Polish National Science Center through the OPUS grant No. 2017/25/B/ST9/01208.
The work of N.\.{Z}. is supported by the South African Research Chairs Initiative (grant no. 64789) of the Department of Science and Innovation and the National Research Foundation\footnote{Any opinion, finding and conclusion or recommendation expressed in this material is that of the authors and the NRF does not accept any liability in this regard.} of South Africa.
J.P.-G. acknowledges funding support from Spanish public funds for research under project ESP2017-87676-C5-5-R and financial support from the State Agency for Research of the Spanish MCIU through the ''Center of Excellence Severo Ochoa'' award to the Instituto de Astrof\'isica de Andaluc\'ia (SEV-2017-0709).
\section*{Data Availability}
The data underlying this article will be shared on reasonable request to the authors.




\bibliographystyle{mnras}
\bibliography{example} 

\begin{thebibliography}{}
\makeatletter
\relax
\def\mn@urlcharsother{\let\do\@makeother \do\$\do\&\do\#\do\^\do\_\do\%\do\~}
\def\mn@doi{\begingroup\mn@urlcharsother \@ifnextchar [ {\mn@doi@}
  {\mn@doi@[]}}
\def\mn@doi@[#1]#2{\def\@tempa{#1}\ifx\@tempa\@empty \href
  {http://dx.doi.org/#2} {doi:#2}\else \href {http://dx.doi.org/#2} {#1}\fi
  \endgroup}
\def\mn@eprint#1#2{\mn@eprint@#1:#2::\@nil}
\def\mn@eprint@arXiv#1{\href {http://arxiv.org/abs/#1} {{\tt arXiv:#1}}}
\def\mn@eprint@dblp#1{\href {http://dblp.uni-trier.de/rec/bibtex/#1.xml}
  {dblp:#1}}
\def\mn@eprint@#1:#2:#3:#4\@nil{\def\@tempa {#1}\def\@tempb {#2}\def\@tempc
  {#3}\ifx \@tempc \@empty \let \@tempc \@tempb \let \@tempb \@tempa \fi \ifx
  \@tempb \@empty \def\@tempb {arXiv}\fi \@ifundefined
  {mn@eprint@\@tempb}{\@tempb:\@tempc}{\expandafter \expandafter \csname
  mn@eprint@\@tempb\endcsname \expandafter{\@tempc}}}

\bibitem[\protect\citeauthoryear{{Abdo} et~al.,}{{Abdo} et~al.}{2010}]{Abdo10}
{Abdo} A.~A.,  et~al., 2010, \apj, 715, 429

\bibitem[\protect\citeauthoryear{{Abdollahi} et~al.,}{{Abdollahi}
  et~al.}{2020}]{Abdollahi}
{Abdollahi} S.,  et~al., 2020, \mn@doi [\apjs] {10.3847/1538-4365/ab6bcb},
  \href {https://ui.adsabs.harvard.edu/abs/2020ApJS..247...33A} {247, 33}

\bibitem[\protect\citeauthoryear{{Akiyama}, {Stawarz}, {Tanaka}, {Nagai},
  {Giroletti}  \& {Honma}}{{Akiyama} et~al.}{2016}]{Akiyama16}
{Akiyama} K.,  {Stawarz} {\L}.,  {Tanaka} Y.~T.,  {Nagai} H.,  {Giroletti} M.,
   {Honma} M.,  2016, \mn@doi [\apjl] {10.3847/2041-8205/823/2/L26}, \href
  {https://ui.adsabs.harvard.edu/abs/2016ApJ...823L..26A} {823, L26}

\bibitem[\protect\citeauthoryear{{Atwood} et~al.,}{{Atwood}
  et~al.}{2009}]{Atwo09}
{Atwood} W.~B.,  et~al., 2009, \mn@doi [\apj] {10.1088/0004-637X/697/2/1071},
  \href {https://ui.adsabs.harvard.edu/abs/2009ApJ...697.1071A} {697, 1071}

\bibitem[\protect\citeauthoryear{{Bachev}, {Mukhopadhyay}  \&
  {Strigachev}}{{Bachev} et~al.}{2015}]{bachev15}
{Bachev} R.,  {Mukhopadhyay} B.,   {Strigachev} A.,  2015, \mn@doi [\aap]
  {10.1051/0004-6361/201425563}, \href
  {https://ui.adsabs.harvard.edu/abs/2015A&A...576A..17B} {576, A17}

\bibitem[\protect\citeauthoryear{{B\"{o}ttcher}, {Harris}  \&
  {Krawczynski}}{{B\"{o}ttcher} et~al.}{2012}]{2012rjag.book.....B}
{B\"{o}ttcher} M.,  {Harris} D.~E.,   {Krawczynski} H.,  2012, {Relativistic
  Jets from Active Galactic Nuclei}

\bibitem[\protect\citeauthoryear{{Cecconi}, {Cencini}  \& {Vulpiani}}{{Cecconi}
  et~al.}{2010}]{Cecconi}
{Cecconi} F.,  {Cencini} M.,   {Vulpiani} A.,  2010, {Chaos: from Simple Models
  to Complex Systems}.
World Scientific, Singapore

\bibitem[\protect\citeauthoryear{{Chen}, {Pohl}, {B{\"o}ttcher}  \&
  {Gao}}{{Chen} et~al.}{2016}]{chen16}
{Chen} X.,  {Pohl} M.,  {B{\"o}ttcher} M.,   {Gao} S.,  2016, \mn@doi [\mnras]
  {10.1093/mnras/stw528}, \href
  {https://ui.adsabs.harvard.edu/abs/2016MNRAS.458.3260C} {458, 3260}

\bibitem[\protect\citeauthoryear{{Costamante} et~al.,}{{Costamante}
  et~al.}{2001}]{Costamante01}
{Costamante} L.,  et~al., 2001, \mn@doi [\aap] {10.1051/0004-6361:20010412},
  \href {https://ui.adsabs.harvard.edu/abs/2001A&A...371..512C} {371, 512}

\bibitem[\protect\citeauthoryear{{Eckmann} \& {Ruelle}}{{Eckmann} \&
  {Ruelle}}{1992}]{eckmann92}
{Eckmann} J.~P.,  {Ruelle} D.,  1992, \mn@doi [Physica D Nonlinear Phenomena]
  {10.1016/0167-2789(92)90023-G}, \href
  {https://ui.adsabs.harvard.edu/abs/1992PhyD...56..185E} {56, 185}

\bibitem[\protect\citeauthoryear{{Finke} \& {Becker}}{{Finke} \&
  {Becker}}{2014}]{finke14}
{Finke} J.~D.,  {Becker} P.~A.,  2014, \mn@doi [\apj]
  {10.1088/0004-637X/791/1/21}, \href
  {https://ui.adsabs.harvard.edu/abs/2014ApJ...791...21F} {791, 21}

\bibitem[\protect\citeauthoryear{{Finke} \& {Becker}}{{Finke} \&
  {Becker}}{2015}]{finke15}
{Finke} J.~D.,  {Becker} P.~A.,  2015, \mn@doi [\apj]
  {10.1088/0004-637X/809/1/85}, \href
  {https://ui.adsabs.harvard.edu/abs/2015ApJ...809...85F} {809, 85}

\bibitem[\protect\citeauthoryear{{Fraser} \& {Swinney}}{{Fraser} \&
  {Swinney}}{1986}]{fraser86}
{Fraser} A.~M.,  {Swinney} H.~L.,  1986, \mn@doi [\pra]
  {10.1103/PhysRevA.33.1134}, \href
  {https://ui.adsabs.harvard.edu/abs/1986PhRvA..33.1134F} {33, 1134}

\bibitem[\protect\citeauthoryear{{Gardner} et~al.,}{{Gardner}
  et~al.}{2006}]{jwst}
{Gardner} J.~P.,  et~al., 2006, \mn@doi [\ssr] {10.1007/s11214-006-8315-7},
  \href {https://ui.adsabs.harvard.edu/abs/2006SSRv..123..485G} {123, 485}

\bibitem[\protect\citeauthoryear{{Grassberger} \& {Procaccia}}{{Grassberger} \&
  {Procaccia}}{1983}]{grassberger83}
{Grassberger} P.,  {Procaccia} I.,  1983, \mn@doi [Physica D Nonlinear
  Phenomena] {10.1016/0167-2789(83)90298-1}, \href
  {https://ui.adsabs.harvard.edu/abs/1983PhyD....9..189G} {9, 189}

\bibitem[\protect\citeauthoryear{{Grassberger}, {Hegger}, {Kantz}, {Schaffrath}
   \& {Schreiber}}{{Grassberger} et~al.}{1993}]{Grassberger}
{Grassberger} P.,  {Hegger} R.,  {Kantz} H.,  {Schaffrath} C.,   {Schreiber}
  T.,  1993, \mn@doi [Chaos] {10.1063/1.165979}, \href
  {https://ui.adsabs.harvard.edu/abs/1993Chaos...3..127G} {3, 127}

\bibitem[\protect\citeauthoryear{{Hanslmeier} et~al.,}{{Hanslmeier}
  et~al.}{2013}]{Hanslmeier}
{Hanslmeier} A.,  et~al., 2013, \mn@doi [A\&A] {10.1051/0004-6361/201015215},
  550, A6

\bibitem[\protect\citeauthoryear{{Hegger} \& {Kantz}}{{Hegger} \&
  {Kantz}}{1999}]{Ka}
{Hegger} R.,  {Kantz} H.,  1999, \mn@doi [\pre] {10.1103/PhysRevE.60.4970},
  \href {https://ui.adsabs.harvard.edu/abs/1999PhRvE..60.4970H} {60, 4970}

\bibitem[\protect\citeauthoryear{{Hegger}, {Kantz}  \& {Schreiber}}{{Hegger}
  et~al.}{1999}]{TISEAN}
{Hegger} R.,  {Kantz} H.,   {Schreiber} T.,  1999, \mn@doi [Chaos]
  {10.1063/1.166424}, \href
  {https://ui.adsabs.harvard.edu/abs/1999Chaos...9..413H} {9, 413}

\bibitem[\protect\citeauthoryear{{Kantz} \& {Schreiber}}{{Kantz} \&
  {Schreiber}}{2004}]{Kantz}
{Kantz} H.,  {Schreiber} T.,  2004, {Nonlinear Time Series Analysis}

\bibitem[\protect\citeauthoryear{{Kennel}, {Brown}  \& {Abarbanel}}{{Kennel}
  et~al.}{1992}]{kennel92}
{Kennel} M.~B.,  {Brown} R.,   {Abarbanel} H. D.~I.,  1992, \mn@doi [\pra]
  {10.1103/PhysRevA.45.3403}, \href
  {https://ui.adsabs.harvard.edu/abs/1992PhRvA..45.3403K} {45, 3403}

\bibitem[\protect\citeauthoryear{{Kidger}, {Gonzalez-Perez}  \&
  {Sadun}}{{Kidger} et~al.}{1996}]{kidger96}
{Kidger} M.~R.,  {Gonzalez-Perez} J.~N.,   {Sadun} A.,  1996, in {Miller}
  H.~R.,  {Webb} J.~R.,   {Noble} J.~C.,  eds,  Astronomical Society of the
  Pacific Conference Series Vol. 110, Blazar Continuum Variability. p.~123

\bibitem[\protect\citeauthoryear{Kodba, Perc  \& Marhl}{Kodba et~al.}{2005}]{U}
Kodba S.,  Perc M.,   Marhl M.,  2005, \mn@doi [EUROPEAN JOURNAL OF PHYSICS
  Eur. J. Phys] {10.1088/0143-0807/26/1/021}, 26, 205

\bibitem[\protect\citeauthoryear{{Lehto}, {Czerny}  \& {McHardy}}{{Lehto}
  et~al.}{1993}]{lehto93}
{Lehto} H.~J.,  {Czerny} B.,   {McHardy} I.~M.,  1993, \mn@doi [\mnras]
  {10.1093/mnras/261.1.125}, \href
  {https://ui.adsabs.harvard.edu/abs/1993MNRAS.261..125L} {261, 125}

\bibitem[\protect\citeauthoryear{{Lichtenberg} \& {Lieberman}}{{Lichtenberg} \&
  {Lieberman}}{1992}]{lichtenberg92}
{Lichtenberg} A.~J.,  {Lieberman} M.~A.,  1992, {Regular and Chaotic Dynamics}.
Springer, New York

\bibitem[\protect\citeauthoryear{{Mandelbrot}}{{Mandelbrot}}{1983}]{mandelbrot83}
{Mandelbrot} B.,  1983, {The Fractal Geometry of Nature}.
W. H. Freeman and Company, New York

\bibitem[\protect\citeauthoryear{{Mannattil}, {Gupta}  \&
  {Chakraborty}}{{Mannattil} et~al.}{2016}]{mannattil16}
{Mannattil} M.,  {Gupta} H.,   {Chakraborty} S.,  2016, \mn@doi [\apj]
  {10.3847/1538-4357/833/2/208}, \href
  {https://ui.adsabs.harvard.edu/abs/2016ApJ...833..208M} {833, 208}

\bibitem[\protect\citeauthoryear{{Misra}, {Harikrishnan}, {Mukhopadhyay},
  {Ambika}  \& {Kembhavi}}{{Misra} et~al.}{2004}]{misra04}
{Misra} R.,  {Harikrishnan} K.~P.,  {Mukhopadhyay} B.,  {Ambika} G.,
  {Kembhavi} A.~K.,  2004, \mn@doi [\apj] {10.1086/421005}, \href
  {https://ui.adsabs.harvard.edu/abs/2004ApJ...609..313M} {609, 313}

\bibitem[\protect\citeauthoryear{Oprisan, Lynn, Tompa  \& Lavin}{Oprisan
  et~al.}{2015}]{Oprisan}
Oprisan S.,  Lynn P.,  Tompa T.,   Lavin A.,  2015, \mn@doi [Frontiers in
  Computational Neuroscience] {10.3389/fncom.2015.00125}, 9, 125

\bibitem[\protect\citeauthoryear{{Padovani}}{{Padovani}}{2017}]{Pado17}
{Padovani} P.,  2017, \mn@doi [Nature Astronomy] {10.1038/s41550-017-0194},
  \href {https://ui.adsabs.harvard.edu/abs/2017NatAs...1E.194P} {1, 0194}

\bibitem[\protect\citeauthoryear{{Pascual-Granado}, {Garrido}  \&
  {Su{\'a}rez}}{{Pascual-Granado} et~al.}{2015}]{PG15}
{Pascual-Granado} J.,  {Garrido} R.,   {Su{\'a}rez} J.~C.,  2015, \mn@doi
  [\aap] {10.1051/0004-6361/201425056}, 575, A78

\bibitem[\protect\citeauthoryear{{Provenzale}, {Vio}  \&
  {Cristiani}}{{Provenzale} et~al.}{1994}]{provenzale94}
{Provenzale} A.,  {Vio} R.,   {Cristiani} S.,  1994, \mn@doi [\apj]
  {10.1086/174267}, \href
  {https://ui.adsabs.harvard.edu/abs/1994ApJ...428..591P} {428, 591}

\bibitem[\protect\citeauthoryear{{Ruelle}}{{Ruelle}}{1990}]{ruelle90}
{Ruelle} D.,  1990, \mn@doi [Proceedings of the Royal Society of London Series
  A] {10.1098/rspa.1990.0010}, \href
  {https://ui.adsabs.harvard.edu/abs/1990RSPSA.427..241R} {427, 241}

\bibitem[\protect\citeauthoryear{{Sadun}}{{Sadun}}{1996}]{sadun96}
{Sadun} A.,  1996, in {Miller} H.~R.,  {Webb} J.~R.,   {Noble} J.~C.,  eds,
  Astronomical Society of the Pacific Conference Series Vol. 110, Blazar
  Continuum Variability. p.~86

\bibitem[\protect\citeauthoryear{{Seymour} \& {Lorimer}}{{Seymour} \&
  {Lorimer}}{2013}]{seymour13}
{Seymour} A.~D.,  {Lorimer} D.~R.,  2013, \mn@doi [\mnras]
  {10.1093/mnras/sts060}, \href
  {https://ui.adsabs.harvard.edu/abs/2013MNRAS.428..983S} {428, 983}

\bibitem[\protect\citeauthoryear{Shishikura}{Shishikura}{1998}]{shishikura98}
Shishikura M.,  1998, \mn@doi [Annals of Mathematics] {10.2307/121009}, 147,
  225

\bibitem[\protect\citeauthoryear{{Smith}, {Mushotzky}, {Boyd}, {Malkan},
  {Howell}  \& {Gelino}}{{Smith} et~al.}{2018}]{smith18}
{Smith} K.~L.,  {Mushotzky} R.~F.,  {Boyd} P.~T.,  {Malkan} M.,  {Howell}
  S.~B.,   {Gelino} D.~M.,  2018, \mn@doi [\apj] {10.3847/1538-4357/aab88d},
  \href {https://ui.adsabs.harvard.edu/abs/2018ApJ...857..141S} {857, 141}

\bibitem[\protect\citeauthoryear{{Stawarz} \& {Petrosian}}{{Stawarz} \&
  {Petrosian}}{2008}]{stawarz08}
{Stawarz} {\L}.,  {Petrosian} V.,  2008, \mn@doi [\apj] {10.1086/588813}, \href
  {https://ui.adsabs.harvard.edu/abs/2008ApJ...681.1725S} {681, 1725}

\bibitem[\protect\citeauthoryear{{Takens}}{{Takens}}{1981}]{Takens}
{Takens} F.,  1981, {Detecting strange attractors in turbulence}.
p.~366, \mn@doi{10.1007/BFb0091924}

\bibitem[\protect\citeauthoryear{{Tarnopolski}}{{Tarnopolski}}{2015}]{tarnopolski15}
{Tarnopolski} M.,  2015, \mn@doi [\apss] {10.1007/s10509-015-2379-3}, \href
  {https://ui.adsabs.harvard.edu/abs/2015Ap&SS.357..160T} {357, 160}

\bibitem[\protect\citeauthoryear{Tarnopolski, {\.{Z}}ywucka, Marchenko  \&
  Pascual-Granado}{Tarnopolski et~al.}{2020}]{tarnopolski20}
Tarnopolski M.,  {\.{Z}}ywucka N.,  Marchenko V.,   Pascual-Granado J.,  2020,
  \mn@doi [The Astrophysical Journal Supplement Series]
  {10.3847/1538-4365/aba2c7}, 250, 1

\bibitem[\protect\citeauthoryear{{Tavecchio}, {Bonnoli}  \&
  {Galanti}}{{Tavecchio} et~al.}{2020}]{tavecchio20}
{Tavecchio} F.,  {Bonnoli} G.,   {Galanti} G.,  2020, \mn@doi [\mnras]
  {10.1093/mnras/staa2055}, \href
  {https://ui.adsabs.harvard.edu/abs/2020MNRAS.497.1294T} {497, 1294}

\bibitem[\protect\citeauthoryear{{The Fermi-LAT collaboration}}{{The Fermi-LAT
  collaboration}}{2020}]{Fermi19}
{The Fermi-LAT collaboration} 2020, \mn@doi [\apjs] {10.3847/1538-4365/ab6bcb},
  \href {https://ui.adsabs.harvard.edu/abs/2020ApJS..247...33A} {247, 33}

\bibitem[\protect\citeauthoryear{{Theiler}, {Eubank}, {Longtin}, {Galdrikian}
  \& {Doyne Farmer}}{{Theiler} et~al.}{1992}]{Theiler}
{Theiler} J.,  {Eubank} S.,  {Longtin} A.,  {Galdrikian} B.,   {Doyne Farmer}
  J.,  1992, \mn@doi [Physica D Nonlinear Phenomena]
  {10.1016/0167-2789(92)90102-S}, \href
  {https://ui.adsabs.harvard.edu/abs/1992PhyD...58...77T} {58, 77}

\bibitem[\protect\citeauthoryear{{Urry} \& {Padovani}}{{Urry} \&
  {Padovani}}{1995}]{Urry95}
{Urry} C.~M.,  {Padovani} P.,  1995, \mn@doi [\pasp] {10.1086/133630}, \href
  {https://ui.adsabs.harvard.edu/abs/1995PASP..107..803U} {107, 803}

\bibitem[\protect\citeauthoryear{{Uttley}, {McHardy}  \& {Vaughan}}{{Uttley}
  et~al.}{2005}]{uttley05}
{Uttley} P.,  {McHardy} I.~M.,   {Vaughan} S.,  2005, \mn@doi [\mnras]
  {10.1111/j.1365-2966.2005.08886.x}, 359, 345

\bibitem[\protect\citeauthoryear{Wernecke, Sándor  \& Gros}{Wernecke
  et~al.}{2019}]{Wernecke19}
Wernecke H.,  Sándor B.,   Gros C.,  2019, \mn@doi [Physics Reports]
  {https://doi.org/10.1016/j.physrep.2019.08.001}, 824, 1

\bibitem[\protect\citeauthoryear{{Wolf}, {Swift}, {Swinney}  \&
  {Vastano}}{{Wolf} et~al.}{1985}]{wolf85}
{Wolf} A.,  {Swift} J.~B.,  {Swinney} H.~L.,   {Vastano} J.~A.,  1985, \mn@doi
  [Physica D Nonlinear Phenomena] {10.1016/0167-2789(85)90011-9}, \href
  {https://ui.adsabs.harvard.edu/abs/1985PhyD...16..285W} {16, 285}

\bibitem[\protect\citeauthoryear{{Wood}, {Caputo}, {Charles}, {Di Mauro},
  {Magill}, {Perkins}  \& {Fermi-LAT Collaboration}}{{Wood}
  et~al.}{2017}]{Wood17}
{Wood} M.,  {Caputo} R.,  {Charles} E.,  {Di Mauro} M.,  {Magill} J.,
  {Perkins} J.~S.,   {Fermi-LAT Collaboration} 2017, in 35th International
  Cosmic Ray Conference (ICRC2017). p.~824 (\mn@eprint {arXiv} {1707.09551})

\makeatother
\end{thebibliography}




\appendix


\bsp	
\label{lastpage}
\end{document}